\documentclass[11pt]{article}
\usepackage{amssymb}
\usepackage{geometry}
\usepackage{amsmath}
\usepackage{amsfonts}
\usepackage{graphicx}

\begin{document}
\newcommand{\2}{\vspace{0.2 cm}}
\newcommand{\dist}{{\rm dist}}
\newcommand{\diam}{{\rm diam}}
\newcommand{\rad}{{\rm rad}}
\newcommand{\dom}{\mbox{$\rightarrow$}}
\newcommand{\ndom}{\mbox{$\not\rightarrow$}}
\newcommand{\sdom}{\mbox{$\Rightarrow$}}
\newcommand{\nsdom}{\mbox{$\not\Rightarrow$}}
\newcommand{\qed}{\hfill$\diamond$}
\newcommand{\pf}{{\bf Proof: }}
\newtheorem{theorem}{Theorem}[section]
\newcommand{\ra}{\rangle}
\newcommand{\la}{\langle}
\newtheorem{lemma}[theorem]{Lemma}
\newtheorem{corollary}[theorem]{Corollary}
\newtheorem{proposition}[theorem]{Proposition}
\newtheorem{conjecture}[theorem]{Conjecture}
\newtheorem{problem}[theorem]{Problem}
\newtheorem{remark}[theorem]{Remark}
\newtheorem{example}[theorem]{Example}
\newcommand{\beq}{\begin{equation}}
\newcommand{\eeq}{\end{equation}}
\newcommand{\argmax}{{\rm argmax}}
\newcommand{\MiP}{MinHOMP($H$) }
\newcommand{\MaP}{MaxHOMP($H$) }
\newcommand{\vecc}[1]{\stackrel{\leftrightarrow}{#1}}

\title{Minimum Cost and List Homomorphisms to Semicomplete Digraphs}
\author{G. Gutin, A. Rafiey and A. Yeo\\
Department of Computer Science\\
Royal Holloway University of London\\
Egham, Surrey TW20 OEX, UK\\
{\tt gutin(arash,anders)@cs.rhul.ac.uk}}

\date{}

\maketitle

\begin{abstract}
For digraphs $D$ and $H$, a mapping $f:\ V(D)\dom V(H)$ is a {\em
homomorphism of $D$ to $H$} if $uv\in A(D)$ implies $f(u)f(v)\in
A(H).$ Let $H$ be a fixed directed or undirected graph. The
homomorphism problem for $H$ asks whether a directed or undirected
graph input digraph $D$ admits a homomorphism to $H.$ The list
homomorphism problem for $H$ is a generalization of the
homomorphism problem for $H$, where every vertex $x\in V(D)$ is
assigned a set $L_x$ of possible colors (vertices of $H$).

The following optimization version of these decision problems was
introduced in \cite{gutinDAM}, where it was motivated by a
real-world problem in defence logistics. Suppose we are given a
pair of digraphs $D,H$ and a positive cost $c_i(u)$ for each $u\in
V(D)$ and $i\in V(H)$. The cost of a homomorphism $f$ of $D$ to
$H$ is $\sum_{u\in V(D)}c_{f(u)}(u)$. For a fixed digraph $H$, the
minimum cost homomorphism problem for $H$, MinHOMP($H$), is stated
as follows: For an input digraph $D$ and costs $c_i(u)$ for each
$u\in V(D)$ and $i\in V(H)$, verify whether there is a
homomorphism of $D$ to $H$ and, if it exists, find such a
homomorphism of minimum cost.

We obtain dichotomy classifications of the computational
complexity of the list homomorphism problem and MinHOMP($H$), when
$H$ is a semicomplete digraph (a digraph in which every two
vertices have at least one arc between them). Our dichotomy for
the list homomorphism problem coincides with the one obtained by
Bang-Jensen, Hell and MacGillivray in 1988 for the homomorphism
problem when $H$ is a semicomplete digraph: both problems are
polynomial solvable if $H$ has at most one cycle; otherwise, both
problems are NP-complete. The dichotomy for \MiP is different: the
problem is polynomial time solvable if $H$ is acyclic or $H$ is a
cycle of length 2 or 3; otherwise, the problem is NP-hard.
\end{abstract}

\section{Introduction}

For excellent introductions to homomorphisms in directed and
undirected graphs, see \cite{hell2003,hell2004}. In this paper,
directed (undirected) graphs have no parallel arcs (edges) or
loops. The vertex (arc) set of a digraph $G$ is denoted by $V(G)$
($A(G)$). The vertex (edge) set of an undirected graph $G$ is
denoted by $V(G)$ ($E(G)$). For a digraph $G$, if $xy\in A(G)$, we
say that $x$ {\em dominates} $y$ and $y$ is {\em dominated} by
$x$. A $k$-{\em cycle}, denoted by $\vec{C}_k$, is a directed
simple cycle with $k$ vertices. A digraph is {\em acyclic} if it
has no cycle. A digraph $D$ is {\em semicomplete} if, for each
pair $x,y$ of distinct vertices either $x$ dominates $y$ or $y$
dominates $x$ or both. A {\em tournament} is a semicomplete
digraph with no 2-cycle. Semicomplete digraphs and, in particular,
tournaments are well-studied in graph theory and algorithms
\cite{bang2000}. A digraph $G'$ is the {\em dual} of a digraph $G$
if $G'$ is obtained from $G$ by changing orientations of all arcs.

For digraphs $D$ and $H$, a mapping $f:\ V(D)\dom V(H)$ is a {\em
homomorphism of $D$ to $H$} if $uv\in A(D)$ implies $f(u)f(v)\in
A(H).$ A homomorphism $f$ of $D$ to $H$ is also called an {\em
$H$-coloring} of $G$, and $f(x)$ is called  {\em color} of $x$ for
every $x\in V(D).$ We denote the set of all homomorphisms from $D$
to $H$ by $HOM(D,H)$. Let $H$ be a fixed digraph. The {\em
homomorphism problem for} $H$, ${\rm HOMP}(H)$, asks whether there
is a homomorphism of an input digraph $D$ to $H$ (i.e., whether
$HOM(D,H)\neq \emptyset$). In the {\em list homomorphism problem
for} $H$, ${\rm LHOMP}(H)$, we given an input digraph $D$ and a
set (called a {\em list}) $L_v\subseteq V(H)$ for each $v\in
V(D)$. Our aim is to check whether there is a homomorphism $f\in
HOM(D,H)$ such that $f(v)\in L_v$ for each $v\in V(D).$

The problems ${\rm HOMP}(H)$ and ${\rm LHOMP}(H)$ have been
studied for several families of directed and undirected graphs
$H$, see, e.g., \cite{hell2003,hell2004}. A well-known result of
Hell and Ne\v{s}et\v{r}il \cite{hellJCT48} asserts that ${\rm
HOMP}(H)$ for undirected graphs is polynomial time solvable if $H$
is  bipartite and it is NP-complete, otherwise. Feder, Hell and
Huang \cite{federC19} proved that ${\rm LHOMP}(H)$ for undirected
graphs is polynomial time solvable if $H$ is a bipartite graph
whose complement is a circular arc graph (a graph isomorphic to
the intersection graph of arcs on a circle), and ${\rm LHOMP}(H)$
is NP-complete, otherwise. Such a dichotomy classification for all
digraphs is unknown and only partial classifications have been
obtained; see \cite{hell2004}. For example, Bang-Jensen, Hell and
MacGillivray \cite{bangSIAMJDM1} showed that ${\rm HOMP}(H)$ for
semicomplete digraphs $H$ is polynomial time solvable if $H$ has
at most one cycle and ${\rm HOMP}(H)$ is NP-complete, otherwise.
Nevertheless, Bulatov \cite{bulatovACMTCL} managed to prove that
for each directed graph $H$, ${\rm LHOMP}(H)$ is either polynomial
time solvable or NP-complete. The same result for ${\rm HOMP}(H)$
is conjectured, see, e.g., \cite{hell2003,hell2004}. If this
conjecture holds, it will imply that the well-known Constraint
Satisfaction Problem Dichotomy Conjecture of Feder and Vardi also
holds \cite{federSIAMJC28}.

The authors of \cite{gutinDAM} introduced an optimization problem
on $H$-colorings for undirected graphs $H$, MinHOMP($H$). The
problem is motivated by a problem in defence logistics. Suppose we
are given a pair of digraphs $D,H$ and a positive cost $c_i(u)$
for each $u\in V(D)$ and $i\in V(H)$. The {\em cost} of a
homomorphism $f$ of $D$ to $H$ is $\sum_{u\in V(D)}c_{f(u)}(u)$.
For a fixed digraph $H$, the {\em minimum cost homomorphism
problem} \MiP is stated as follows: For an input digraph $D$ and
costs $c_i(u)$ for each $u\in V(D)$ and $i\in V(H)$, verify
whether $HOM(D,H)\neq \emptyset$ and, if $HOM(D,H)\neq \emptyset$,
find a homomorphism in $HOM(D,H)$ of minimum cost. The problem
\MiP generalizes ${\rm LHOMP}(H)$ (and, thus, HOMP($H$)): assign
$c_i(u)=1$ if $i\in L_u$ and $c_i(u)=2,$ otherwise.

In this paper, we obtain dichotomy classifications for ${\rm
LHOMP}(H)$ and \MiP when $H$ is a semicomplete digraph. Our
classification for ${\rm LHOMP}(H)$ coincides with that for ${\rm
HOMP}(H)$ \cite{bangSIAMJDM1} described earlier. However, for \MiP
the classification is different: the problem is polynomial time
solvable when $H$ is either an acyclic tournament or a 2-cycle or
a 3-cycle. Otherwise, \MiP is NP-hard. This implies that even when
$H$ is a unicyclic semicomplete digraph on at least four vertices,
\MiP is NP-hard (unlike HOMP($H$) and LHOMP($H$)).

The {\em maximum cost homomorphism problem} \MaP is the same
problem as MinHOMP($H$), but instead of minimization we consider
maximization. Let $M$ be a constant larger than any cost $c_i(u)$,
$u\in V(D),\ i\in V(H)$. Then the cost $c'_i(u)=M-c_i(u)$ is
positive for each $u\in V(D),\ i\in V(H)$. Due to this
transformation, the problems \MiP and \MaP are equivalent. Notice
that allowing negative or zero costs would not make \MiP and \MaP
more difficult: we can easily transform this more general case to
the positive costs one by adding a large constant $M'$ to each
cost. This transformation does not change optimal solutions.

The rest of the paper is organized as follows. In Section
\ref{homsec}, we introduce the homomorphic product of digraphs and
relate it to a maximum cost homomorphism. We prove that LHOMP($H$)
and \MiP are polynomial time solvable when $H$ is an acyclic
tournament. The dichotomy classifications LHOMP($H$) and \MiP when
$H$ is a semicomplete digraph are proved in Sections \ref{lhclass}
and \ref{semsec}, respectively. We conclude the paper by posing
some open problems.

\section{Products and Homomorphisms of Digraphs}\label{homsec}

In this section, we describe an approach for proving that \MaP is
polynomial time solvable for certain digraphs $H$. Using our
approach, we prove that \MaP is polynomial time solvable for
acyclic tournaments. To the best of our knowledge this approach,
which is of interest also for ${\rm HOM}(H)$ and ${\rm LHOM}(H)$,
has not been studied earlier.

For $H$ belonging to a special family $\cal H$ of digraphs, we can
transform \MaP into the problem of finding a maximum cost
independent set in a special family ${\cal F}({\cal H})$ of
undirected graphs. If the last problem is polynomial time solvable
(when, for example, ${\cal F}({\cal H})$ consists of perfect
graphs, $2P_2$-free graphs, claw-free graphs or graphs of other
special classes, see
\cite{alekseev1991,alekseevDAM145,balasN19,bertolazziDAM76,grotschelADM21,nakamuraJORSJ44}),
then our approach is useful.

The {\em homomorphic product}  of digraphs $D$ and $H$ is an
undirected graph $D\otimes H$ defined as follows: $V(D\otimes
H)=\{u_i:\ u\in V(D),\ i\in V(H)\}$, $E(D\otimes H)=\{u_iv_j:\
uv\in A(D),\ ij\notin A(H)\}\cup \{u_iu_j: u\in V(D), i\neq j\in
V(H)\}.$ Let $\mu=\max\{c_j(v):\ v\in V(D),\ j\in V(H)\}.$ We
define the cost of $u_i$, $c(u_i)=c_i(u)+\mu |V(D)|.$ For a set
$X\subseteq V(D\otimes H)$, we define $c(X)=\sum_{x\in X}c(x).$

\begin{theorem}\label{maint}
Let $D$ and $H$ be digraphs. Then there is a homomorphism of $D$
to $H$ if and only if the number of vertices in a largest
independent set of $D\otimes H$ equals $|V(D)|$. If $HOM(D,H)\neq
\emptyset$, then $h\in HOM(D,H)$ is of maximum cost if and only if
$I=\{x_{h(x)}:\ x\in V(D)\}$ is an independent set of maximum
cost.
\end{theorem}
\pf Let $h:\ D \dom H$ be a homomorphism. Consider $I=\{x_{h(x)}:\
x\in V(D)\}$. Suppose that $x_{h(x)}y_{h(y)}$ is an edge in
$D\otimes H$. Then either $xy\in A(D)$ and $h(x)h(y)\notin A(H)$
or $yx\in A(D)$ and $h(y)h(x)\notin A(H)$. Either case contradicts
the fact that $h$ is a homomorphism. Thus, $I$ is an independent
set in $D\otimes H.$

Observe that each independent set in $D\otimes H$ contains at most
one vertex in each set $S_x=\{x_i:\ i\in V(H)\},$ $x\in V(D)$. Let
$I=\{x_{f(x)}:\ x\in V(D)\}$ be an independent set in $D\otimes H$
with $|V(D)|$ vertices. Consider the mapping $f:\ x\mapsto f(x).$
Assume $xy\in A(D).$ Since $I$ is independent, $f(x)f(y)\in A(H)$.
Thus, $f\in HOM(D,H).$

Let $HOM(D,H)\neq \emptyset$ and let $n=|V(D)|.$ Let $X$ and $Y$
be subsets of $V(D\otimes H)$ and $|X|=|Y|+1\le n.$ Then
$$ c(X)-c(Y)\ge |X|n\mu -(|X|-1)(n+1)\mu\ge \mu
>0.$$
Thus, in particular, every maximum cost independent set of
$D\otimes H$ is a largest independent set. Observe that the cost
of the homomorphism $f$ defined above equals the cost of vertices
in the independent set $I$ minus $n^2\mu,$ which is a constant.
Thus, every maximum cost independent set of $D\otimes H$
corresponds to a maximum cost homomorphism of $D$ to $H$ and vise
versa.\qed

\begin{remark}
Observe that Theorem \ref{maint} holds when $H$ has loops.
\end{remark}

\begin{remark}
In applications of Theorem \ref{maint}, we may need to replace a
pair $D,H$ by another pair $D',H'$ such that $HOM(D,H)=HOM(D',H')$
and the costs of the homomorphisms remain the same.
\end{remark}

Consider the following corollary of Theorem \ref{maint}. A digraph
$D$ is {\em transitive} if $xy,yz\in A(D)$ implies $xz\in A(D)$
for all pairs $xy,yz$ of arcs in $D.$ A graph is a {\em
comparability graph} if it has an orientation, which is
transitive. Bang-Jensen, Hell and MacGillivray \cite{bangSIAMJDM1}
proved that if $H$ is an acyclic tournament, then ${\rm HOMP}(H)$
is polynomial time solvable. We extend this result to \MiP and
MaxHOMP($H$).

\begin{theorem}\label{acyclicth}
If $H$ is an acyclic tournament, then \MaP and \MiP are polynomial
time solvable.
\end{theorem}
\pf Let $H$ be an acyclic tournament with $V(H)=\{1,2,\ldots,p\}$
and $A(H)=\{ij: 1\le i<j\le p\}.$

Observe that $H$ is transitive. Also observe that $HOM(D,H)=
\emptyset$ unless $D$ is acyclic. Since we can verify that $D$ is
acyclic in polynomial time (for example, by deleting vertices of
indegree 0), we may assume that $D$ is acyclic. Since $H$ is
transitive, we have $HOM(D,H)=HOM(D^+,H)$, where $D^+$ is the
transitive closure of $D$, i.e., if there is a path from $x$ to
$y$ in $D$, then $xy\in D^+.$ One can find the transitive closure
of a digraph in polynomial time using DFS or BFS \cite{bang2000},
so we may assume that $D$ is transitive.

Let $G=D\otimes H$. Let $G'$ be an orientation of $G$ such that
$$A(G')=\{x_iy_j:\ j\leq i, xy\in A(D)\}\cup \{x_ix_j:\ x\in
V(D),\ j<i\}.$$ We will prove that $G'$ is a transitive digraph.
Let $x_iy_j,y_jz_k\in A(G')$. Observe that $i\ge j\ge k$ and
consider three cases covering all possibilities.

\2

{\bf Case 1:} $x=y=z$. Then $x_ix_j,x_jx_k\in A(G')$ and, thus,
$i>j>k$ and $x_iz_k=x_ix_k \in A(G').$

\2

{\bf Case 2:}  $x=y=z$ does not hold, but not all vertices $x,y,z$
are distinct. Without loss of generality, assume that $x=y\neq z.$
Then $x_ix_j,x_jz_k\in A(G')$ and, thus, $i>k$ and $x_iz_k\in
A(G').$

\2

{\bf Case 3:} $x,y,z$ are all distinct. Then $xy,yz\in A(D^+)$
and, thus, $xz\in A(D^+)$. Since $i\ge k$, we conclude that
$x_iz_k\in A(G').$

\2

So, we have proved that $G$ is a comparability graph and, thus, it
is perfect. Therefore, a maximum cost independent set in $D\otimes
H$ can be found in polynomial time \cite{grotschelADM21}. It
remains to apply Theorem \ref{maint}. If $D\otimes H$ has an
independent set with $|V(D)|$ vertices, $HOM(D,H)\neq \emptyset$
and a maximum cost independent set corresponds to a maximum cost
$H$-coloring.\qed

\begin{corollary}
If $H$ is an acyclic tournament, then LHOMP($H$) is polynomial
time solvable.
\end{corollary}

\section{Dichotomy for LHOMP($H$)}\label{lhclass}

Recall that $\vec{C}_k$ denotes a directed cycle on $k$ vertices,
$k\ge 2$; let $V(\vec{C}_k)=\{1,2,\ldots,k\}.$  One can check
whether $HOM(D,\vec{C}_k)\neq \emptyset$ using the following
algorithm $\cal A$ from Section 1.4 of \cite{hell2004}. First, we
may assume that $D$ is connected (i.e., its underlying undirected
graph is connected) as otherwise $\cal A$ can be applied to each
component of $D$ separately. Choose a vertex $x$ of $D$ and assign
it color 1. Assign every out-neighbor of $x$ color 2 and each
in-neighbor of $x$ color $k$. For every vertex $y$ with color $i$,
we assign every out-neighbor of $y$ color $i+1$ modulo $k$ and
every in-neighbor of $y$ color $i-1$ modulo $k$. We have
$HOM(D,\vec{C}_k)\neq \emptyset$ if and only if no vertex is
assigned different colors.

M. Green \cite{green} was the first to prove Theorem \ref{P3}, but
his proof uses polymorphisms (for the definition and results on
polymorphisms, see, e.g., \cite{bulatovACMTCL}). Our proof below
is elementary and does not require polymorphisms.

\begin{theorem}\label{P3}
Let $H$ be a semicomplete digraph with a unique cycle, then ${\rm
LHOMP}(H)$ is polynomial time solvable.
\end{theorem}
\pf It is well-known \cite{bang2000} that a semicomplete digraph
with a unique cycle contains a cycle with two or three vertices.
Assume that $H$ has a cycle with three vertices (the case of
2-cycle can be treated similarly). Let the vertex set of $H$ be
$\{1,2,\ldots,p\}$ and $A(H)=\{ij:\ i<j, (i,j)\neq (a,b)\}\cup
\{ba\}$, where $b=a+2.$

We use the following recursive procedure. If $V(D)= \{v\}$ and
$L_v \neq \emptyset $, then the solution is trivial. Now suppose
that $|V(D)| \ge 2$ and consider the following two properties of a
vertex $x$ in $D$:

\begin{itemize}
\item[(a)] $x$ has in-degree zero, and $L_x$ has an element
smaller than $a$. \item[(b)] $x$ has out-degree zero, and $L_x$
has an element greater than
  $a+2$.
\end{itemize}

If there is a vertex $x$ with property $(a)$, then define
$f(x)=i$, where $i$ is the minimum number in $L_x$, and delete all
$j \le i$ from the lists of all out-neighbors of $x$. Run the
procedure for $D-x$ with changed lists.  If there is a vertex $x$
with property $(b)$, then define $f(x)=i$, where $i$ is the
maximum number in $L_x$, and delete all $j \ge i$ from the lists
of all in-neighbors of $x$. Run the procedure for $D-x$ with
changed lists.

If no vertex with either property exists, then run the algorithm
$\cal A$ described in the beginning of this section to find all
homomorphisms from $HOM(D,\vec{C}_3),$ where $\vec{C}_3$ has
vertices $a,a+1,a+2$. If $HOM(D,\vec{C}_3)\neq \emptyset$, there
there are three homomorphism, and it suffices to verify that at
least one of them is compatible with the lists.

Clearly, if our procedure succeeds, then we have found a required
homomorphism. It remains to see that if the procedure fails, then
no required homomorphism exists. This is equivalent to proving
that, after all vertices satisfying (a) or (b) have been deleted,
every remaining vertex $y$ must have color $a,a+1$ or $a+2.$

Let $D'$ be obtained from $D$ by deleting all vertices satisfying
(a) or (b) and let $y\in V(D').$ We prove that $y$ must be colored
$a,a+1$ or $a+2.$ Assume that $y$ is in a directed cycle $C$ of
$D'$. Observe that any homomorphism of $D$ to $H$ maps $C$ into a
directed walk. Thus, $y$ can be colored $a,a+1$ or $a+2$ only.
Assume that $y$ is isolated in $D'$. Since $y$ does not satisfy
(a) or (b), its list contains only $a$, $a+1$ or $a+2.$ Now
consider the case when $y$ is not isolated and it is not in any
cycle of $D'.$ Let $P$ be a path in $D'$ containing $y$ such that
the initial vertex of $P$ is either in a cycle of $D'$ or its
in-degree in $D'$ is zero, and the terminal vertex of $P$ is
either in a cycle of $D'$ or its out-degree in $D'$ is zero.
Observe that, by the arguments above, the initial vertex of $P$
must have color $a$ or larger and the terminal vertex of $P$ must
have color $a+2$ or smaller. This implies that every vertex of $P$
must have color $a$, $a+1$ or $a+2.$ \qed

\2

Recall that HOMP($H$) is NP-complete when $H$ is a semicomplete
digraph with at least two cycles. This result and Theorems
\ref{acyclicth} and \ref{P3} imply the following:

\begin{theorem}
Let $H$ be a semicomplete digraph. Then LHOMP($H$) is polynomial
time solvable if $H$ has at most one cycle, and LHOMP($H$) is
NP-complete, otherwise.
\end{theorem}

\section{Classification for \MiP and \MaP}\label{semsec}

To solve \MiP for $H=\vec{C}_k$, choose an initial vertex $x$ in
each component $D'$ of $D$ (a component of its underlying
undirected graph). Using the algorithm $\cal A$ from the previous
section, we can check whether each $D'$ admits an
$\vec{C}_k$-coloring. If the coloring of $D'$ exists, we compute
the cost of this coloring and compute the costs of the other $k-1$
$\vec{C}_k$-colorings when $x$ is colored $2,3,\ldots,k,$
respectively. Thus, we can find a minimum cost homomorphism in
$HOM(D',\vec{C}_k).$  Thus, in polynomial time, we can obtain a
$\vec{C}_k$-coloring of the whole digraph $D$ of minimum cost. In
other words, we have the following:

\begin{lemma}\label{cycle}
For $H=\vec{C}_k$, \MiP and \MaP are polynomial time solvable.
\end{lemma}

Addition of an extra vertex to a cycle may well change the
complexity of \MaP and MinHOMP($H$).

\begin{lemma}\label{cycle+}
Let $H'$ be a digraph obtained from $\vec{C}_k$, $k\ge 2$, by
adding an extra vertex dominated by the vertices of the cycle, and
let $H$ be $H'$ or its dual. Then \MiP and \MaP are NP-hard.
\end{lemma}
\pf Without loss of generality we may assume that $H=H'$ and that
$V(H)=\{1,2,3,\ldots , k, k+1\}$, $123 \ldots k1$ is a $k$-cycle,
and the vertex $k+1$ is dominated by the vertices of the cycle.

We will reduce the maximum independent set problem to
MinHOMP($H$). Let $G$ be a graph. Construct a digraph $D$ as
follows:

$$V(D)=V(G) \cup \{v_i^e :\ e\in E(G) \ i\in V(H) \},\
A(D)=A_1\cup A_2, \mbox{ where}$$
$$A_1=\{v_1^e v_2^e, v_2^ev_3^e, \ldots v_{k-1}^ev_k^e,v_k^ev_1^e :\ e\in
E(G)\}$$ and
$$A_2=\{v_1^{uv}u, v_{k+1}^{uv}u, v_2^{uv}v, v_{k+1}^{uv}v :\ uv\in E(G)\}.$$

Let all costs $c_i(t)=1$ for $t\in V(D)$ apart from $c_{k+1}(p)=2$
for all $p\in V(G).$

Consider a minimum cost homomorphism $f\in HOM(D,H)$. By the
choice of the costs, $f$ assigns the maximum possible number of
vertices of $G$ (in $D$) a color different from $k+1$. However, if
$pq$ is an edge in $G$, by the definition of $D$, $f$ cannot
assign colors different from $k+1$ to both $p$ and $q$. Indeed, if
both $p$ and $q$ are assigned colors different from $k+1$, then
the existence of $v_{k+1}^{pq}$ implies that they are assigned the
same color, which however is impossible by the existence of
$\{v_i^{pq} :\  i\in \{1,2,\ldots ,k\} \}$. Observe that $f$ may
assign exactly one of the vertices $p,q$ color $k+1$ and the other
a color different from $k+1$. Also $f$ may assign both of them
color $k+1$. Thus, a minimum cost $H$-coloring of $D$ corresponds
to a maximum independent set in $G$ and vise versa (the vertices
of a maximum independent set are assigned color $2$ and all other
vertices in $V(G)$ are assigned color $k+1$). \qed \2

Interestingly, the problem ${\rm HOMP}(H')$  for $H'$ (especially,
with $k=3$) defined in Lemma \ref{cycle+} is well known to be
polynomial time solvable (see, e.g.,
\cite{bangSIAMJDM1,gutjahr1991,hell2004}). The following lemma
allows us to prove that \MaP and \MiP are NP-hard when
MaxHOMP($H'$) and MinHOMP($H'$) are NP-hard for an induced
subdigraph $H'$ of $H.$

\begin{lemma}\label{reductionl}
Let $H'$ be an induced subdigraph of a digraph $H$. If
MaxHOMP($H'$) is NP-hard, then \MaP is also NP-hard.
\end{lemma}
\pf Let $D$ be an input digraph with $n$ vertices and let $c_i(u)$
be the costs, $u\in V(D)$, $i\in V(H')$. Let all costs $c_i(u)$ be
bounded from above by $\beta(n)$. For each $i\in V(H)-V(H')$ and
each $u\in V(D)$, set costs $c_i(u):=n\beta(n)+1.$ Observe that
there is an $H$-coloring of $D$ of cost at most $n\beta(n)$ if and
only if $HOM(D,H')\neq \emptyset$ and if $HOM(D,H')\neq
\emptyset$, then the cost of minimum cost $H$-coloring equals to
that of minimum cost $H'$-coloring.\qed

\2

As a corollary of Theorem \ref{acyclicth} and Lemmas \ref{cycle},
\ref{cycle+} and \ref{reductionl}, we obtain the following
theorem.

\begin{theorem}\label{classmd}
For a semicomplete digraph $H$, \MiP and \MaP are polynomial time
solvable if $H$ is acyclic or $H=\vec{C}_k$ for $k=2$ or 3, and
NP-hard, otherwise.
\end{theorem}
\pf By Theorem \ref{acyclicth} and since ${\rm HOMP}(H)$ is
NP-complete when a semicomplete digraph $H$ has at least two
cycles \cite{bangSIAMJDM1}, we may restrict ourselves to the case
when $H$ has a unique cycle. Observe that this cycle has two or
three vertices. If no other vertices are in $H$, \MaP and \MiP are
polynomial time solvable by Lemma \ref{cycle}. Assume that $H$ has
a vertex $i$ not in the cycle. Observe that $i$ is dominated by or
dominates all vertices of the cycle, i.e., $H$ contains, as an
induced subdigraph one of the digraphs of Lemma \ref{cycle+}. So,
we are done by Lemmas \ref{cycle+} and \ref{reductionl}.\qed

\section{Discussions}

In this paper we have obtained dichotomy classifications for the
list and minimum cost $H$-coloring problems when $H$ is a
semicomplete digraph.

It would be interesting to find out whether there exists a
dichotomy classification for the minimum cost $H$-coloring problem
(for an arbitrary digraph $H$) and if it does exist, to obtain
such a classification. Since these problems seem to be far from
trivial, one could concentrate on establishing dichotomy
classifications for special classes of digraphs such as
semicomplete multipartite digraphs (digraphs obtained from
complete multipartite graphs by replacing every edge with an arc
or the pair of mutually opposite arcs) and oriented paths (it was
proved in \cite{gutjahrDAM35} that the homomorphism problem for
oriented paths is polynomial time solvable).

We have obtained some partial results on MinHOMP($H$) for
semicomplete multipartite digraphs $H$. To find a complete
dichotomy for the case of semicomplete bipartite digraphs, one
would need, among other things, to solve an open problem from
\cite{gutinDAM}: establish a dichotomy classification for the
complexity of \MiP when $H$ is a bipartite (undirected) graph.
Indeed, let $B$ be a semicomplete bipartite digraph with partite
sets $U,V$ and arc set $A(B)=A_1\cup A_2$, where $A_1=U\times V$
and $A_2\subseteq V\times U$. Let $B'$ be a bipartite graph with
partite sets $U,V$ and edge set $E(B')=\{uv:\ vu\in A_2\}.$
Observe that MinHOMP($B$) is equivalent to MinHOMP($B'$).

It was proved in \cite{gutinDAM} that \MiP is polynomial time
solvable when $H$ is a bipartite graph whose complement is an
interval graph. It follows from the main result of \cite{federC19}
that \MiP is NP-hard when $H$ is a bipartite graph whose
complement is not a circular arc graph. This leaves the obvious
gap in the classification for \MiP when $H$ is a bipartite graph.

\2

\2

{\bf Acknowledgement} We are thankful to Dave Cohen, Mike Green,
Michael Krivelevich, Alek Vainshtein and others for useful
discussions on the topic of the paper.

\end{document}